\documentclass[twocolumn]{autart}
\usepackage{amsmath,amsfonts,amssymb,graphicx,stmaryrd,multirow,color}
\newcommand{\Hi}{\mathcal{H}_\infty}
\newcommand{\be}{\begin{equation}}
\newcommand{\ee}{\end{equation}}
\newcommand{\bea}{\begin{eqnarray}}
\newcommand{\eea}{\end{eqnarray}}
\newtheorem{assumption}{Assumption}

\newcommand{\R}{\mathbb{R}}
\newcommand{\RR}{\mathbb{R}}

\newcommand{\C}{\mathbb{C}}
\newcommand{\Z}{\mathbb{Z}}

\newcommand{\w}{\omega}
\newcommand{\A}{\mathbb A}

\begin{document}
\begin{frontmatter}

\title{Computation of Extremum Singular Values and \\ the Strong H-infinity Norm of SISO Time-Delay Systems}

\author[First]{Suat Gumussoy},
\author[Second]{Wim Michiels}

\address[First]{MathWorks, \\
        3 Apple Hill Drive, Natick, MA 01760, USA \\
        \mbox{(e-mail: suat.gumussoy@mathworks.com)}.}
\address[Second]{Department of Computer Science, KU Leuven, \\
        Celestijnenlaan 200A, 3001, Heverlee, Belgium \\
        \mbox{(e-mail: wim.michiels@cs.kuleuven.be)}.}

\begin{abstract}
We consider the computation of H-infinity norms for Single-Input-Single-Output (SISO) time-delay systems, which are described by delay differential algebraic equations. Unlike the iterative level set methods in the literature, we present a novel numerical method to compute the H-infinity norm. This method requires solving one eigenvalue problem of at most twice the size of the eigenvalue problem in every iteration of a level set method, but in practice often considerably lower. We first show that the computation of extrema of the transfer function can be turned into the computation of the imaginary axis zeros of a transcendental function. We compute these zeros by a predictor-corrector type algorithm.
It is known that the H-infinity norm of delay differential algebraic systems, which can model both retarded and neutral type systems, might be sensitive with respect to arbitrarily small delay perturbations. This recently led to the concept of strong H-infinity norms, which explicitly take into account such small delay perturbations. We present a direct numerical method to compute the strong H-infinity norm of SISO time-delay systems.  Our algorithm is applicable to the closed-loop system of interconnections (series, parallel, feedback, junctions) of time-delay systems and/or controllers.
\end{abstract}
\end{frontmatter}
\section{Introduction}
The availability of robust methods to compute $\Hi$ norms is essential in a computer aided control system design \cite{ZhouRobustBook1995}. The common approaches for computing $\Hi$ norms for finite-dimensional linear systems, which belong to the class of level set methods, are based on the relation between the intersections of singular value curves of the transfer function with a constant function (the level) and the presence of imaginary axis eigenvalues of a corresponding Hamiltonian matrix  \cite{ByersSISC1988}.  In \cite{ByersSISC1988,BoydMCSS1989}  linearly converging bisection based algorithms have been proposed. Quadratically convergent algorithms, relying on an alternating search in two directions, have been described in \cite{BoydSCL1990,BruinsmaSCL1990}. In \cite{MichielsSIMAX2010} a level set algorithm for computing $\Hi$ norms of a class of retarded type time-delay systems has been outlined. The main complication with respect to the delay-free case is that the intersections between the singular value curves and a level set are no longer related to the spectrum of a Hamiltonian matrix, but to an infinite-dimensional operator.
%
%
 Therefore, the algorithm of \cite{MichielsSIMAX2010} adopts a predictor-corrector approach, where in the first step the operator is discretized using a spectral method, followed by local corrections to remove the effect of the discretization on computed peak values in the frequency response.   For the sake of completeness, it should be mentioned that for systems without delay, $\Hi$ norms can also be computed by a parameter sweep, thereby checking the feasibility of linear matrix inequalities, see, e.g.,~\cite{scherer}. However, since these methods implicitly construct a Lyapunov function, and in the infinite-dimensional,  time-delay case, fixing the form of the Lyapunov functional to a tractable form involving finitely many free parameters typically introduces conservatism, only potentially conservative bounds on the $\Hi$ norm can be aimed at in the latter case.

Inherent to level set methods, whether for systems with or without delay, is that the $\Hi$ norm is computed in an iterative way, by updating the level in every iteration step until convergence to the dominant peak in the singular value plot is achieved. In every iteration step the imaginary axis eigenvalues of a Hamiltonian matrix or operator need to be computed. As a first main contribution of the paper, we present a novel numerical algorithm to compute $\Hi$ norms of SISO time-delay systems requiring solving only \emph{one} eigenvalue problem.

The second main contribution is that our method admits a system description in a \emph{standard form}, described by a set of delay differential algebraic equations (DDAEs).  As we will see, this form contains a large set of systems, including interconnections of time-delay systems and controllers in complex configurations, and including both retarded and neutral type systems.

Recently, we analyzed in \cite{GumussoySICON2011} the properties of the $\Hi$ norm of time-delay systems. We illustrated that the $\Hi$ norm of DDAEs may be sensitive with respect to arbitrarily small delay perturbations. Due to this sensitivity, we introduced the {\em strong $\Hi$ norm,} which explicitly takes into account small delay perturbations, inevitable in any practical control application, and we outlined the computation using a level set approach. The derived theory of strong $\Hi$ norms can be considered as the dual of the theory of strong stability for neutral systems and DDAEs, as, e.g., elaborated in \cite{have:02,wimdae} and the references therein. As a third, main contribution, the presented algorithm in this paper takes the potential sensitivity problem into account, by computing the strong $\Hi$ norm. In control problems without feed through at infinity along the loops, this strong $\Hi$ norms reduces to the standard $\Hi$ norm.

The remainder of the paper organizes as follows. Section~\ref{sec:transform} describes the standard form on time-delay systems on which the algorithms rely. Section \ref{sec:extremumsv} presents the computation of extrema of the transfer function of a SISO time-delay system. In particular, \S\ref{sec:extremumsv} shows that the local maximum and minimum can be computed by finding the imaginary axis zeros of a transcendental function. In \S\ref{sec:zeroscomp} these zeros are computed by a predictor-corrector algorithm. Based on this computation and and a characterization of the high-frequency behavior, we present a non-iterative numerical algorithm for the strong $\Hi$ norm computation in Section~\ref{sec:shinfnorm}. Numerical examples and concluding remarks are given in Sections~\ref{sec:ex} and \ref{sec:conc}.

\noindent \textbf{Notation:} The sets of complex, real and integer numbers are $\C, \R, \Z$ and the set of  strictly positive and nonnegative real numbers are $\R^+,\R_0^+$. Zero and identity matrices are $0, I$. The transpose of the matrix $A$ is $A^T$. Complex conjugate transpose, i\textsuperscript{th} maximum singular value and derivative of transfer function $G(s)$ are shown as $G^*(s)$, $\sigma_i(G)$, $G^{\prime}(s)$ respectively. The function constructing block diagonal matrix from input arguments is \verb"blkdiag".
\section{Standard form for time-delay systems} \label{sec:transform}
%
%

We consider a general representation for a time-delay system described by the delay differential algebraic equations,
\be \label{eq:dae}
G:\left\{
  \begin{array}{ll}
    E\dot{\bar{x}}(t)=A_0 \bar{x}(t)+\sum_{i=1}^m A_{i}\bar{x}(t-\tau_i)+Bu(t), \\
    y(t)=C\bar{x}(t),
  \end{array}
\right.
\ee where $E$, $A_i\in\R^{n \times n}$ for $i=0,\ldots,m$, $B\in\R^{n \times n_u}$ and $C\in\R^{n_y \times n}$ are real valued matrices with appropriate dimensions. The time-delays $\tau_i$ for $i=0,\ldots,m$ are non-negative real numbers.
Let matrix $E$ in (\ref{eq:dae}) satisfy
\[
\mathrm{rank}(E)=n-\nu,
\]
with $0\leq \nu<n$. In case $\nu\geq 1$, i.e., matrix $E$ is singular, we let the columns
of matrix $U\in\RR^{n\times \nu}$, respectively $V\in\RR^{n\times \nu}$, be a (minimal) orthonormal basis for
the left, respectively right nullspace of $E$, which implies $U^T E=0$, $E V=0$.
We then make the following assumption.
\begin{assumption} \label{assumptionddae1}
Matrix $U^T A_0 V$ is nonsingular.
\end{assumption}
This assumption is necessary to make system (\ref{eq:dae}) causal~\cite{wimdae}. We refer to this reference for a discussion of basic properties of (\ref{eq:dae})(definition of solutions, spectrum determined growth property of solutions,\ldots).

\begin{assumption} \label{assumptionstability}
The null solution of system (\ref{eq:dae}), with $u\equiv 0$, is strongly exponentially stable.
\end{assumption}
The stability of the time-delay system (\ref{eq:dae}) is a necessary assumption for $\Hi$ norm computation since this norm is finite for stable systems only. Strong exponential stability refers to the fact that the asymptotic stability of the null solution is robust against small delay perturbations, \cite{have:02}.

The representation in (\ref{eq:dae}) is the standard form of the SISO time-delay plant for our algorithms. This form is \emph{closed} under block diagram operations such as series, parallel, feedback etc. and therefore, it is rich enough to represent most of the systems as a result of interconnections of time-delay systems and/or controllers. This form (\ref{eq:dae}) allows to represent systems with multiple state, multiple input and output delays, as well as systems with a nontrivial feed through including neutral type systems, see \cite{GumussoySICON2011}.
This generality is an important property since the $\Hi$ norm computation of the \emph{closed-loop} system with the designed controller is a common scenario. Therefore representing the closed-loop system and all its subsystems with the same form is essential for practical use of the algorithm.

\begin{figure}[htbp]
\begin{center}
\includegraphics[width=0.49\textwidth]{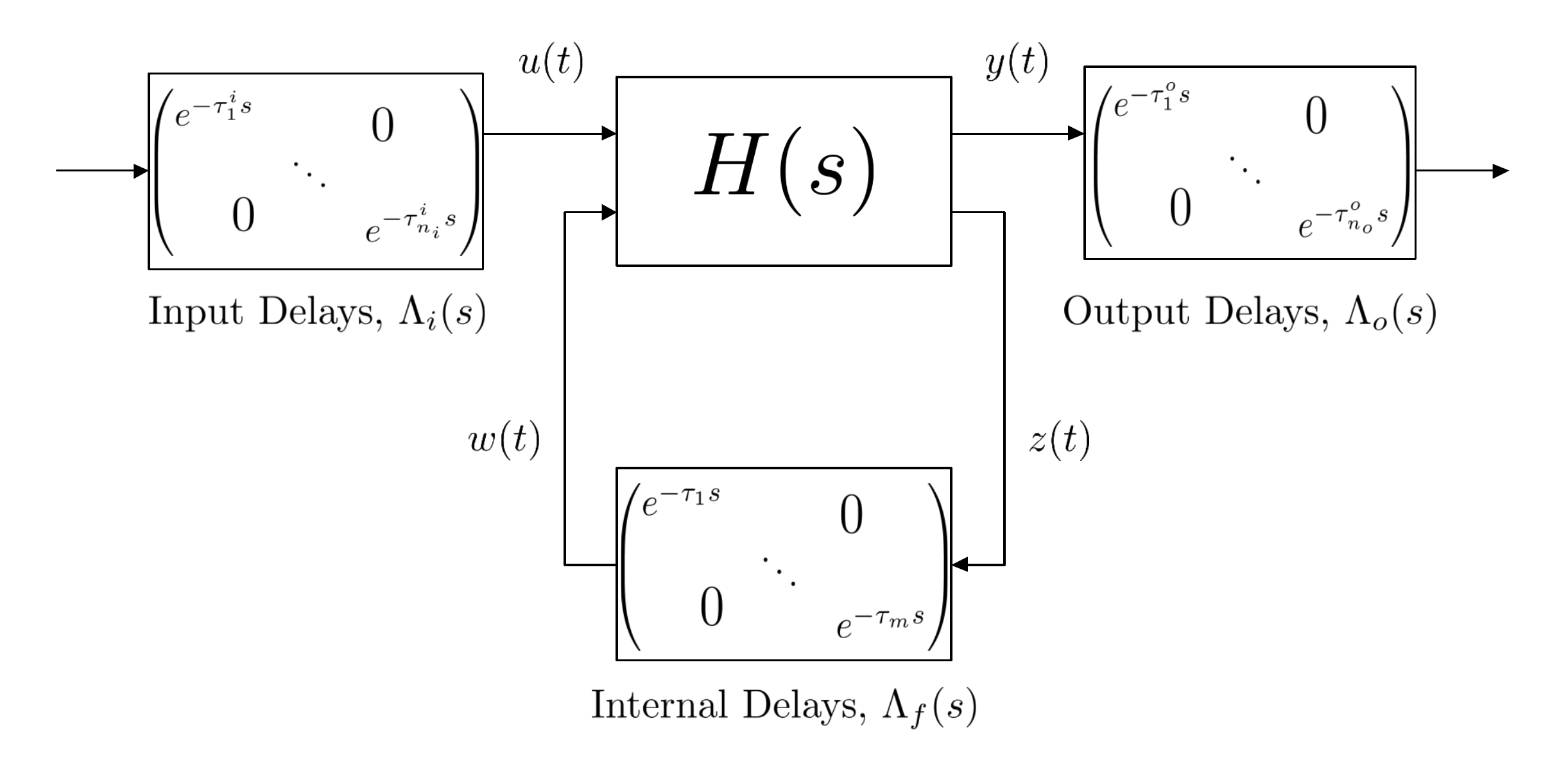}
\caption{Systems with internal, input and output delays} \label{fig:tmwdelaymodel}
\end{center}
\end{figure}

Another representation for the time-delay systems (also used in MATLAB) is the \emph{generalized LTI} (GLTI) class  of continuous-time LTI systems, where systems are modeled as the LFT interconnection of a delay-free LTI model $H$ and a set of internal, input and output delays (see Figure \ref{fig:tmwdelaymodel}).

The class of GLTI systems is also closed under series, parallel, and feedback connections as well as branching/summing junctions \cite{GahinetACC2004}.

State-space equations for $H(s)$ and time-delay terms in Figure~\ref{fig:tmwdelaymodel} are
\bea
F\dot{x}(t) & = & A x(t) + B_1 u(t) + B_2 w(t) \label{eq:Hs} \\
\nonumber y(t) & = & C_1 x(t) + D_{11} u(t) + D_{12} w(t) \\
\nonumber z(t) & = & C_2 x(t) + D_{21} u(t) + D_{22} w(t) \\
\nonumber w(t) & = & (\Lambda_f z) (t)
\eea
where $\Lambda_f z$ is the vector-valued signal defined by
$(\Lambda_f z) (t) := (z_1^T (t - \tau_1), \hdots, z_m^T (t - \tau_m))^T$. $\Lambda_i$ and $\Lambda_o$ operate on signals similarly, see \cite{GumussoyIFACTDS2012} for further details on defining various types of systems with time-delays.
We can transform the time-delay system in LFT form (\ref{eq:Hs}) into our standard form (\ref{eq:dae}) by first defining the augmented state $\bar{x}^T:=[x^T\ \gamma_u^T\ w^T\ z^T]^T$ where $\gamma_u$ is the auxiliary variable for the input signal $u$. Then we rewrite system equations (\ref{eq:Hs}) in terms of the new state variable $\bar{x}$ as in (\ref{eq:dae})
\bea
\nonumber E&=&\verb"blkdiag"(F,0,0,0), \\ 
\nonumber A_0&=&\left[ 
  \begin{array}{cccc}
    A & B_1 & B_2 & 0 \\
    C_2 & D_{21} & D_{22} & -I \\
    0 & 0 & -I & 0 \\
    0 & -I & 0 & 0 \\
  \end{array}
\right], A_i=\left[ 
  \begin{array}{cccc}
    0 & 0 & 0 & 0 \\
    0 & 0 & 0 & 0 \\
    0 & 0 & 0 & I \\
    0 & 0 & 0 & 0 \\
  \end{array}
\right] \\
\nonumber B&=&\left[ 
  \begin{array}{cccc}
    0 & 0 & 0 & I\\
  \end{array}
\right]^T, C=
\left[
\begin{array}{cccc}
C_1 & D_{11} & D_{12} & 0
\end{array}
\right].
\eea
Hence, the transformation of  (\ref{eq:Hs}) to (\ref{eq:dae}) is immediate.

In the remainder of the paper we assume that, $n_u=n_y=1$,\ i.e.,~(\ref{eq:dae}) is a SISO system.
\section{Computation of extremum singular values} \label{sec:extremumsv}
%
The transfer function of (\ref{eq:dae}) is equal to
\be \label{eq:Glong}
G(s)=C(sE-A_0-\sum_{i=1}^m A_i e^{-s\tau_i})^{-1} B.
\ee
In what follows we characterize local extrema of the function
\begin{equation}\label{extrema}
\R \ni\omega\mapsto |G(j\omega)|.
\end{equation}

An extremum of the singular value curve of $G(j\omega)$ satisfies
\[
\begin{array}{lll}
0 &= &\frac{d}{d\omega} G(j\omega) G(j\omega)^*=\frac{d}{d\omega} G(j\omega) G(-j\omega)\\
 &=& j G^{\prime}(j\omega)G(-j\omega)-j G(j\omega)G^{\prime}(-j\omega).
\end{array}
\]
We arrive at the following result.
\begin{prop}
 The extrema of (\ref{extrema}) and corresponding frequencies can be obtained from the imaginary axis zeros of the transcendental function
\begin{equation} \label{eq:GGp}
Z(s):=G^\prime(s)G(-s)-G^\prime(-s)G(s).
\end{equation}
\end{prop}

We introduce the following notation and two lemmas to derive a state-space like representation  (\ref{eq:GGp}). Let $\A(s)=A_0+\sum_{i=1}^m A_i e^{-s\tau_i}$ and let $\A'$ be the derivative of $\A$ with respect to $s$. Then, we can write the transfer function of the SISO time-delay system (\ref{eq:Glong}) as
\be \label{eq:G}
G(s)=C(sE-\A(s))^{-1}B.
\ee

\begin{lem} \label{lem:Gp}
Let $G(s)$ have a transfer function (\ref{eq:G}).  The derivative of $G(s)$ with respect to $s$ is
\[
G^\prime(s)= \left[C\ \ 0\right]
\left(
s \left[\begin{array}{cc} E & 0 \\ 0 & E  \end{array}\right]
-\left[\begin{array}{cc} \A & \A'-E \\ 0 & \A  \end{array}\right]
\right)^{-1}
\left[\begin{array}{c} 0 \\ B \end{array}\right].
\]
\end{lem}
\noindent\textbf{Proof.} A simple computation yields
\bea
\nonumber G'(s)&=&-C(s E -\A)^{-1}(E-\A')(sE -\A)^{-1}B, \\
\nonumber &=& \left[C\ \ 0\right]
\left[\begin{array}{cc} sE-\A & E-\A' \\ 0 & sE-\A  \end{array}\right]^{-1}
\left[\begin{array}{c} 0 \\ B \end{array}\right].\hspace*{1.5cm} \begin{array}{r} \\ \Box\end{array}
\eea

Using Lemma~\ref{lem:Gp}, we can derive the state space representation for $G'(s)$ and $G'(-s)$.  The state space representation of (\ref{eq:GGp}) can be derived by the following result, which are natural extension of finite dimensional case.
\begin{lem} \label{lem:AddMul}
Let $G_1(s)=C_1(sE_1-\A_1)^{-1}B_1$ and $G_2(s)=C_2(sE_2-\A_2)^{-1}B_2$. Then
\[
G_1(s)+G_2(s)=
[C_1\ C_2]
\left[\begin{array}{cc}
sE_1-\A_1 & 0 \\
0 & sE_2-\A_2
\end{array}\right]^{-1}
\left[\begin{array}{c}
B_1 \\
B_2
\end{array}\right]
\]
and
\[
G_2(s) G_1(s)= [0\ C_2]
\left[\begin{array}{cc}
sE_1-\A_1 & 0 \\
-B_2 C_1 & sE-\A_2
\end{array}\right]^{-1}
\left[\begin{array}{l}
B_1 \\
0
\end{array}\right].
\]
\end{lem}

By Lemma~\ref{lem:Gp} and \ref{lem:AddMul}, we arrive at the following representation of $Z(s)$ in (\ref{eq:GGp}):
\begin{prop}
We can express
 \begin{equation}\label{Zss}
 Z(s)=C_z(s E_z-\A_z(s))^{-1}B_z,
 \end{equation}
  where the matrices are given by
\bea
\nonumber E_z&=&\verb"blkdiag"(-E,E,E,-E,E,E), \\
\nonumber \A_z(s)&=&\verb"blkdiag"(\mathcal{A}_z(s),\mathcal{A}_z(-s)), \\
\nonumber B_z&=& \left[ \begin{array}{cccccc} B^T & 0 & 0 & B^T & 0 & 0 \end{array} \right]^T,
 C_z=\left[ \begin{array}{cccccc} 0 & C & 0 & 0 & C & 0 \end{array} \right],
\eea
and
$\mathcal{A}_z(s)=
    \left[
    \begin{array}{ccc}
      \A(-s) & 0  & 0        \\
       0   & \A(s) & \A'(s)-E \\
       BC  & 0 &   \A(s)
    \end{array}
    \right]\in\mathbb{R}^{3n\times 3n}.
$
\end{prop}

 The next section presents the numerical algorithm to compute the zeros of $Z$ using the above representation.

\subsection{Computing the zeros of $Z$} \label{sec:zeroscomp}
Based on (\ref{Zss}) we can compute the imaginary axis zeros of $Z$ in (\ref{eq:GGp}) by computing the imaginary axis solutions of  the following nonlinear eigenvalue problem
\begin{equation} \label{eq:TEVP}
\begin{array}{ll}
\left[\begin{array}{cc}
\A_z(s)-s E_z & \quad B_z\\
C_z & \quad 0
\end{array}\right]
\end{array}\left[
    \begin{array}{c}
    u \\
    v
    \end{array}
\right]=0,
\end{equation}
where $[u^T,v^T]^T$ is the corresponding eigenvector.

The nonlinearity of the eigenvalue problem stems from the fact that $\A_z(s)$ depends on $e^{-s\tau_i}$ and $e^{s\tau_i},\ i=1,\ldots,m$. This makes that the number of solutions of (\ref{eq:TEVP}) is in general infinite.  It is important to note that the solutions are symmetric with respect to the imaginary axis, as can easily be seen from (\ref{eq:GGp}). Therefore, solutions either appear in quadruples $(s,\bar s,-s,-\bar s)$, or in pairs on the imaginary axis, the latter corresponding to the extrema of the original transfer function.

\medskip
Due to the nonlinearity of (\ref{eq:TEVP}), an approximation  is necessary to globally detect zeros on the imaginary axis. This brings us to a predictor-corrector approach to solve the problem, inspired by~\cite{GumussoySICON2011}.
For the predictor step, we start from a rational approximation of (\ref{eq:Glong}):
\begin{equation} \label{finite2}
G_N(\lambda):={\bf C_N}(\lambda {\bf E_N}-{\bf A_N})^{-1}{\bf B_N},
\end{equation}
obtained by a spectral discretization of the delay system on a grid of $N$ Chebyshev points \cite{BredaANM2006,GumussoySICON2011}. See Appendix B in \cite{GumussoySICON2011} for the computation of the system matrices in (\ref{finite2}). Subsequently, we determine the extrema of the curve $\omega\mapsto G_N(j\omega)$. Similarly to the above derivation, these are given by the imaginary axis zeros of
\begin{equation}\label{defZN}
Z_N(s):=G_N^\prime(s)G_N(-s)-G_N^\prime(-s)G_N(s).
\end{equation} Let $G_N$ has the zero-pole-gain form as $G_N(s)=k\frac{b(s)}{a(s)}$. Then, we can write $Z_N$ as
\begin{multline}
\nonumber Z_N(s)=\overbrace{\left(-\frac{b'(-s)}{b(-s)}+\frac{b'(s)}{b(s)}+\frac{a'(-s)}{a(-s)}-\frac{a'(s)}{a(s)}\right)}^{\Delta(s)}\\ \frac{k^2}{a^3(s)a^3(-s)b(s)b(-s)}.
\end{multline}
Note that the imaginary axis zeros as $Z_N$ and $\Delta$ are the same and we can compute the system $\Delta$ as
\[
\Delta(s)=\sum_{i=1}^{n_z} \left(\frac{1}{s+\hat{z}^*_i}+\frac{1}{s-\hat{z}_i}\right)-\sum_{k=1}^{n_p} \left(\frac{1}{s+\hat{p}^*_k}+\frac{1}{s-\hat{p}_k}\right)
\] where $\hat{z}_i$ and $\hat{p}_k$ are zeros and poles of the system $G_N$. The computation of the imaginary axis zeros of $\Delta$ requires to solve the (standard) generalized eigenvalue problem of size $2(n_p+n_z)+1$,
\begin{equation}\label{disc1}
\begin{array}{ll}
\left[\begin{array}{cc}
A_\Delta-s I & \quad B_\Delta\\
C_\Delta & \quad 0
\end{array}\right]
\end{array}\left[
    \begin{array}{c}
    u_N \\
    v_N
    \end{array}
\right]=0,
\end{equation}
where $A_\Delta$, $B_\Delta$ and $C_\Delta$ are system matrices of the system $\Delta$. Due to the singularity of the matrix $E$ in interconnected system, the computational cost could be significantly lower than the order of system $G_N$, $n(N+1)$.


Because of the error induced by replacing $G(s)$ with $G_N(s)$, the imaginary axis eigenvalues of (\ref{disc1}) will only be approximations of the imaginary axis eigenvalues of (\ref{eq:TEVP}) looked for. Therefore, the second, correction step of the algorithm serves to remove the discretization error. It is based on solving a system of nonlinear equations that characterize extremum points in the singular value curve of $G$, where the initial values are obtained in the first, predictor step. These equations are given by
\begin{equation} \label{eq:Tcorrection}
H(j\omega,\xi)\left[
    \begin{array}{c}
    u \\
    v
    \end{array}
\right]
=0, \quad \Im \{v^* (E_z-\A_z'(j\w))u \}=0,
\end{equation} and the normalization constraints $n(u,v)=0$
where
\[
H(j\omega,\xi):=\left[\begin{array}{cc}
j\omega E-\A(j\omega) & -\xi^{-1}BB^T\\
\xi^{-1}C^TC & j\omega E^T+\A^T(-j\w)
\end{array}\right].
\]
The first equation in (\ref{eq:Tcorrection}) expresses that $\xi$ is a singular value of $G$ at frequency $\omega$ by rewriting the singular value equation in the form of a Hamiltonian eigenvalue problem. The second equation  expresses that the Hamiltonian eigenvalue problem has a double imaginary axis solution. This property corresponds to $\xi$ being an extremum of the singular value curve and is equivalent to a zero derivative of the singular value curve with respect to the $\omega$. Finally normalization constraints for the singular vectors need to be added to make the solution unique. For more details we refer to~\cite{MichielsSIMAX2010}.
%
%
%
\begin{rem}
The reason for solving (\ref{eq:Tcorrection}) in the correction step instead of  solving (\ref{eq:TEVP}) is that the number of equations is smaller ($4n+3$ instead of $12n+2$ real equations).
\end{rem}
The overall algorithm for the computation of extremum singular values is as follows.
\begin{alg} \label{alg:extremum}
 {\em
 \noindent
  Input: system data, $N$, grid $\Omega_N$ with grid points (see \cite{GumussoySICON2011}.)}
\begin{enumerate}
\item \underline{\emph{Prediction step:}} \\
\begin{enumerate}
\item Compute all imaginary axis zeros of $Z_N$ in (\ref{defZN}) by computing the generalized eigenvalues of the pencil (\ref{disc1}),
            whose imaginary axis eigenvalues are given by $\lambda=j\w^{(i)}$.
\item For each imaginary axis eigenvalue, compute the predicted extremum point $\tilde\xi^{(i)}$ of the singular value curve of $G$ in (\ref{eq:dae}) by $\tilde\xi^{(i)}=|G(j\tilde\omega^{(i)})|$.
\end{enumerate}
\item \underline{\emph{Correction step:}} \\
Solve the nonlinear equations (\ref{eq:Tcorrection})
using the Gauss-Newton method,
with the starting values
\[
\omega=\tilde\omega^{(i)},\
\left[\begin{array}{c}u\\
v\end{array}\right]=
\arg\min_{\zeta}{\|H(j\tilde\omega^{(i)},\tilde\xi^{(i)})\zeta\|}/{\|\zeta\|};
 \]
denote the solutions with $(\hat u^{(i)},\hat
 v^{(i)},\hat\omega^{(i)},\hat\xi^{(i)})$, for $i=1,2,....$
\item The extremum singular values of the time-delay system $G$ in (\ref{eq:dae}) and their frequencies are $(\hat\xi^{(i)},\hat\omega^{(i)})$ for $i=1,2,...$.
\end{enumerate}
\end{alg}

There is a linear relationship with the number $N$ and the length of the frequency range approximated. The cut-off frequency as a function of $N$ is illustrated in Figure~4.1 of \cite{MichielsSIMAX2010}. Experience from extensive benchmarking learns that in most practical problems a very small value of $N$ can be taken (the default $N$= 20 is largely sufficient). For further details of the choice of the number of discretization points, $N$, we refer to~\cite{MichielsSIMAX2010}, whose approach extends to the problem considered.

\section{Direct computation of the strong $\Hi$ norm} \label{sec:shinfnorm}
%
The $\Hi$ norm of an asymptotically stable SISO system with transfer function (\ref{eq:Glong}) satisfies
\begin{equation} \label{eq:Ghinf}
||G||_\infty:= \sup_{\omega\geq 0} |G(j\omega)|.
\end{equation}
Algorithm~\ref{alg:extremum} computes finite extrema of the transfer function. This is not sufficient for the following reasons.
\begin{enumerate}
\item Description (\ref{eq:dae}) allows to model systems with a non-trivial feed through. As a consequence, the $\Hi$ norm might not be reached at a finite frequency.
\item It is shown in \cite{MichielsSIMAX2010} that the standard $\Hi$ norm of  linear delay-differential algebraic systems (and neutral type systems)  might be sensitive to infinitesimal perturbations of  the time-delays. The sensitivity  takes, for instance, place in control loops which have a feed trough at infinity, prone to time-delays. It is due to the high frequency behavior of the transfer function  (\ref{eq:Glong}), which is described by the \emph{asymptotic transfer function} $G_a(s)$ defined as
\begin{equation} \label{eq:Ga}
 -C V (U^T A_0 V +\sum_{i=1}^m U^T A_i V e^{-s\tau_i})^{-1} U^TB.
\end{equation}
\end{enumerate}
The above two observations have led in \cite{GumussoySICON2011} to the introduction of the concept of  \emph{strong} $\mathcal{H}_{\infty}$ norm, which is the smallest upper bound robust against infinitesimal delay perturbations.  Making the dependence of $G$ on the delays $\vec\tau=(\tau_1,\ldots,\tau_m)$ explicit with the notation $G(\lambda;\ \vec\tau)$, we have:

\begin{defn}
For given delays $\vec \tau\in(\RR_{0}^+)^{m}$, the strong $\mathcal{H}_{\infty}$ norm of $G$, $\interleave G\interleave_{\infty}$, is defined as
\[
\lim_{\epsilon\rightarrow 0+}\sup \left\{\|G(j\w;\ \vec \tau_\epsilon)\|_{\infty}:\ \
\vec \tau_\epsilon\in\mathcal{B}(\vec \tau,\epsilon) \cap (\RR^+)^{m} \right\}.
\] where $\mathcal{B}(\vec \tau,\epsilon)$ is open ball of radius $\epsilon$ centered at $\vec\tau\in(\R^+)^m$,  $\mathcal{B}(\vec \tau,\epsilon):=\{\vec\theta\in(\R)^m : \|\vec\theta-\vec \tau\|<\epsilon\}$.
\end{defn}
The definition of $\interleave G_a \interleave_{\infty}$ is analogous.   The following results can be found in \cite{GumussoySICON2011}.
\begin{prop}
The assertions below hold.
\begin{itemize}
\item The asymptotic transfer function satisfies
\begin{multline}\label{formstrong}
\interleave G_a\interleave_{\infty}
= \max_{\vec\theta\in [0,\ 2\pi]^m} |C V (U^T A_0 V  \\
+\sum_{i=1}^m U^T A_i V e^{j \theta_i})^{-1} U^TB|
\end{multline} with the argument of the $\max$ operator  continuous in $\vec\theta$.
\item The strong $\Hi$ norm of $G$ is equal to
\begin{equation} \label{eq:Gshinfnorm}
    \interleave G\interleave_\infty=\max\left(\|G\|_{\infty}, \interleave G_a\interleave_\infty \right).
\end{equation}
\end{itemize}
\end{prop}
The first assertion allows a computation of $\interleave G_a\interleave_{\infty}$ by gridding in the $\vec\theta$ space. It should be stressed that in most applications the number of actual time-delays appearing in $G_a$ is much smaller than the number of system delays, $m$, reducing significantly the computational cost. This is because most of the terms in the parenthesis of (\ref{eq:Ga}) are typically zero. The nonzero terms correspond to a high frequency feed through paths over the control loop.

From the second assertion it follows that if $\interleave G\interleave_{\infty}> \interleave G_a\interleave_{\infty}$, the (strong) $\Hi$ norm of $G$ is reached at a finite frequency. Hence, a combination of  computing $\interleave G_a\interleave_{\infty}$ with Algorithm~\ref{alg:extremum} allows to compute $\interleave G\interleave_{\infty}$. We arrive at Algorithm~\ref{alg:hinfnorm}.

\begin{alg} \label{alg:hinfnorm}
 {\em
 \noindent
  Input: system data.}
\begin{enumerate}
\item \underline{\emph{The strong $\Hi$ norm of $G_a$:}} \\ Compute $\xi_s^a$, the strong $\Hi$ norm of the asymptotic transfer function $G_a$, by gridding.
\item \underline{\emph{The strong $\Hi$ norm of $G$:}} \\ By Algorithm~\ref{alg:extremum}, compute $(\xi_o,\w_o)$, the maximum of  $|G(j\omega)|$ in (\ref{eq:dae}) and its frequency, where  $\xi_o=\hat\xi^{(i_o)}=\max( \hat\xi^{(1)},\hat\xi^{(2)},\ldots )$ and $\w_o = \hat\omega^{(i_o)}$.
\\
Compute $\xi_s$, the strong $\Hi$ norm of $G$, where $\xi_s=\max(\xi_s^a,\xi_o)$, and its frequency $\w_s$, where $\w_s=\w_o$ if $\xi_o>\xi_s$ or $\w_s=\infty$ otherwise.
\end{enumerate}
\end{alg}
There are two parts in the computation cost of the strong $\Hi$ norm. The first part is to find the strong $\Hi$ norm of the asymptotic transfer function $G_a$. As pointed out the number of delays appearing in $G_a$ is usually much smaller than the number of system delays.
 Therefore the computation cost for the first step is not usually high. The second main part is the computation of the generalized eigenvalues of  pencil (\ref{disc1}) in the prediction step of Algorithm \ref{alg:extremum} with dimensions $2(n_p+n_z)+1$, where $n_p$ and $n_z$ are the poles and zeros of the descriptor system $G_N$ with order $n(N+1)$. The default value for $N$ is $20$ in our code. Inherent to the DAE modeling framework, matrix $E$ is often singular, leading values of $n_z$ and $n_p$  considerably lower than the order $G_N$ (see the next section for an example). The algorithm for $\Hi$ norm computation in \cite{MichielsSIMAX2010}  only applies to retarded time-delay systems. The $\Hi$ norm computation in \cite{MichielsSIMAX2010} is iterative due to the level set approach and requires solving an eigenvalue problem of size $2n(N+1)$ for each level set. On the other hand, the algorithm in \cite{GumussoySICON2011} considers  retarded and neutral type time-delay systems and has the same numerical cost as Algorithm~2 in the first part and in the second part it solves a generalized eigenvalue problem with dimensions $2n(N+1)$ in every iterative step.


\section{Numerical Examples} \label{sec:ex}

We consider the Smith Predictor example in \cite{GumussoyIFACTDS2012} where the subsystems are $P=6e^{-106s}/(37s+1)$, $G_p=5.6/(40.2s+1)$, $D_p=e^{-93.9s}$, $C=0.5(1+1/(40*s))$ and $F=1/(20s+1)$.

\begin{figure}[!h]
    \begin{minipage}[t]{0.49\textwidth}
        \vspace{0pt}
        \includegraphics[width=\linewidth]{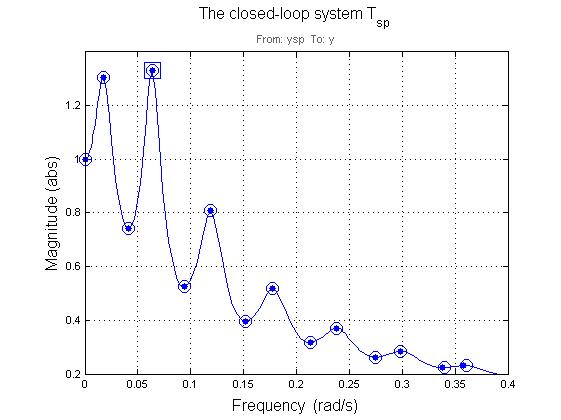}
    \end{minipage}
    \begin{minipage}[t]{0.49\textwidth}
        \vspace{0pt}\raggedright
        \includegraphics[width=\linewidth]{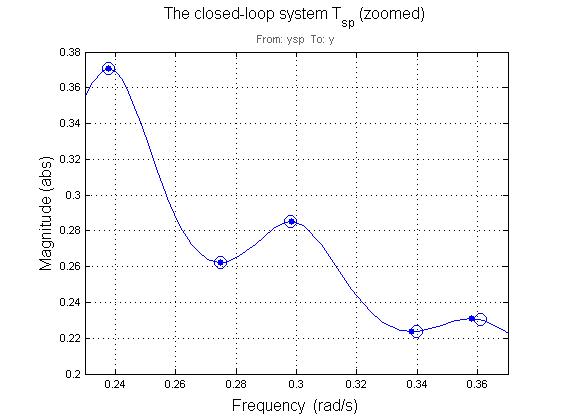}
   \end{minipage}
   \caption{\label{fig:SmithBode} (top) The magnitude Bode plot of $T_{sp}$ and its extremum points computed at prediction and correction steps (bottom), the zoomed version with  last five extremum points.}
\end{figure}

 The closed-loop system $T_{sp}$ is a generalized LTI with internal delays and its Bode magnitude plot is shown in Figure~\ref{fig:SmithBode}. Circles indicate the computed imaginary axis zeros of $Z(s)$ (\ref{eq:GGp}) in the prediction step, for $N=20$. Dots show the results after the prediction results are corrected, inducing a move to the extremum locations of the singular value curve.
Note that the predicted extremum points do not improve much after correction step for initial points since the approximation in the prediction step reliably computes the points. As shown in Figure~\ref{fig:SmithBode} at the bottom, zoomed to larger frequencies, we see the improvement in the correction step due to slight deviation of the approximation from the exact values. The closed-loop system is a retarded time-delay system and its asymptotic transfer function is equal to zero which can be seen from the high frequency behavior in Figure~\ref{fig:SmithBode}. Therefore, the strong $\Hi$ norm of $T_{sp}$ is equal to the standard $\Hi$ norm of $T_{sp}$ by (\ref{eq:Gshinfnorm}). The norm is equal to the largest singular value $1.3308$ of $4^{th}$ point in the extremum points. This example has a singular $E$ matrix, therefore, the algorithm in \cite{MichielsSIMAX2010} is not applicable. When we compared the prediction steps of our algorithm (including pole, zero computation of $G_N$) and the one in \cite{GumussoySICON2011}, the computation time is $0.1885$ and $0.6447$ seconds on Intel Xeon $3.06$GHz with $12$GB RAM. The size of the (single) eigenvalue problem for our algorithm is $175$, whereas the size of the eigenvalue problem in \cite{GumussoySICON2011} is $2n(N+1)=378$ (to be solved in every iteration).

%

\begin{figure}[!h]
    \begin{minipage}[t]{0.49\textwidth}
        \vspace{0pt}
        \includegraphics[width=\linewidth]{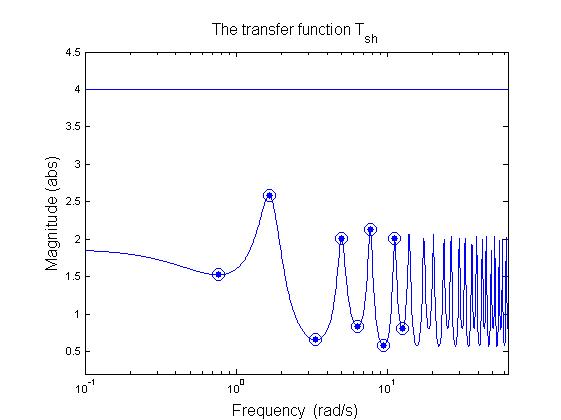}
    \end{minipage}
    \begin{minipage}[t]{0.49\textwidth}
        \vspace{0pt}\raggedright
        \includegraphics[width=\linewidth]{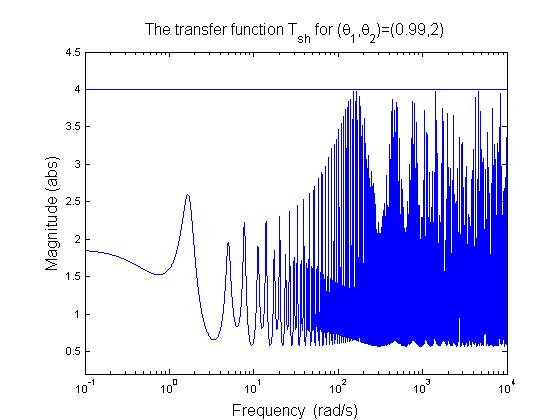}
   \end{minipage}
   \caption{\label{fig:StrongHinf} (top) The magnitude Bode plot of $T_{sh}$ and its extremum points computed at prediction and correction steps (bottom), same Bode plot for $(\tau_1,\tau_2)=(0.99,2)$.}
\end{figure}

\noindent The second example considers the case where the strong $\Hi$ norm,  computed using Algorithm~\ref{alg:hinfnorm}, is different from the standard $\Hi$ norm. Given $\tau_1=1$, $\tau_2=2$, the transfer function $T_{sh}$ is equal to
\[
T_{sh}=(s+2.1) / ((s+0.1)(1-0.25e^{-\tau_1s}+0.5e^{-\tau_2s})+1).
\]
The magnitude plot of $T_{sh}$ is shown in Figure~\ref{fig:StrongHinf} on the top and the extremum points are marked as above.
Note that the high frequency behavior in Figure~\ref{fig:StrongHinf} on the top does not converge to zero. Therefore, the asymptotic transfer function is different from zero. As a first step of the algorithm, the strong $\Hi$ norm of the asymptotic transfer function $T_{a}$, given by
\[
T_{a}=1/({1-0.25e^{-\tau_1s}+0.5e^{-\tau_2s}}),
\]
is computed and equals $4$, obtained  at  $\theta_1=0$ and $\theta_2=\pi$ in formula (\ref{formstrong}).

In the second step, we compute the extrema of the magnitude plot by Algorithm~\ref{alg:extremum}.
The standard $\Hi$ norm of $T_{sh}$ is equal to $2.5788$ and the high frequency behavior visualized in Figure~\ref{fig:StrongHinf} on the top. Therefore, the strong $\Hi$ norm of $T_{sh}$ is equal to $4$ by the final step of Algorithm~\ref{alg:hinfnorm}, which is larger than the standard $\Hi$ norm. This  illustrates that the $\Hi$ norm may be sensitive to small delay changes. Figure~\ref{fig:StrongHinf} at the bottom shows that the strong $\Hi$ norm value $4$ is achieved for a slight perturbation in the delay $\tau_1$ and it can be shown that this norm is attained a larger frequency for smaller delay perturbation size.

\section{Concluding Remarks} \label{sec:conc}

We presented novel algorithms for the computation of extremal singular values and strong $\Hi$ norms of SISO time-delay systems described by DDAEs. The latter algorithm does need the iteration inherent to level set methods.
%

The approach can be easily extended to multi-input-single-output (MISO) and single-input-multiple-output (SIMO) systems. The algorithm namely relies on the property that the singular value plot of the transfer function involves one singular value curve. Whether the presented approach can be extended to general MIMO systems is an open problem.
\section*{Acknowledgements}
This work has been supported by  the Belgian
Federal Science Policy Office, the KU Leuven research council and
 the Research Foundation-Flanders (FWO). The first author thanks Elfin D. Gumussoy for the fruitful discussions.

\bibliographystyle{plain}

\end{document}